\documentclass[11pt,a4paper]{article}

\usepackage{amsmath,amssymb,amsfonts,bbm}
\usepackage[top=3.11cm,bottom=4.11cm,right=3.11cm,left=3.11cm]{geometry}
\usepackage{physics}
\usepackage{slashed}
\usepackage{xcolor,graphicx}
\usepackage{subfigure}
\usepackage{etoolbox}
\usepackage{bbold}
\usepackage{xspace}
\usepackage{appendix}

\usepackage[utf8]{inputenc}
\usepackage[T1]{fontenc}
\usepackage[UKenglish]{babel}

\usepackage[style=ieee, citestyle=numeric-comp, backend=biber]{biblatex}
\numberwithin{equation}{section}
\usepackage{hyperref} 
\usepackage[capitalise]{cleveref}

\usepackage{tabularx}
\newcolumntype{C}{>{\centering\arraybackslash}X}
\newcolumntype{R}{>{\raggedleft\arraybackslash}X}
\newcolumntype{L}{>{\raggedright\arraybackslash}X}

\usepackage{multirow, bigdelim}

\allowdisplaybreaks
\newcommand{\intx}{\int d^4x}
\newcommand{\Dintx}{\int d^Dx}

\title{Full three-loop Renormalisation of an abelian chiral Gauge Theory with non-anticommuting $\gamma_5$ in the BMHV Scheme}

\newcommand{\email}{}
\author{
Dominik  Stöckinger,$^a$\thanks{\email{Dominik.Stoeckinger@tu-dresden.de}}\quad
Matthias
Weißwange,$^a$\thanks{\email{Matthias.Weisswange@tu-dresden.de}}\\[2em]
$^a$Institut für Kern- und Teilchenphysik, TU Dresden,\\ Zellescher Weg 19, DE-01069 Dresden, Germany}

\begin{document}
\thispagestyle{empty}

\maketitle

\setcounter{footnote}{0}
\vspace{2ex}
\begin{abstract}
In this work we present a complete three-loop renormalisation of an abelian chiral gauge 
theory within the Breitenlohner-Maison/'t Hooft-Veltman (BMHV) scheme of dimensional regularisation (DReg). 
In this scheme the 
$\gamma_5$-matrix appearing in gauge interactions is a non-anticommuting object, leading to a breaking 
of gauge and BRST invariance.
Employing an efficient method based on the quantum action principle, we obtain the complete three-loop
counterterm action which serves both
to render the theory finite and to restore gauge and BRST invariance. 
The UV singular counterterms involve not only higher order $\epsilon$-poles 
but also new counterterm structures emerging at the three-loop level for the first time;
the finite symmetry-restoring counterterms are restricted to the same structures as at lower loop orders, just
with different coefficients, aligning with our expectations. Both the singular and the finite counterterms
include structures which cannot be obtained by the standard multiplicative renormalisation.
Our results demonstrate that a rigorous treatment of chiral gauge theories with $\gamma_5$ defined 
in the BMHV scheme at the 
multi-loop level is possible and that the obtained counterterm action is suitable
for computer implementations, allowing automated calculations without ambiguities caused by $\gamma_5$. 
\end{abstract}

\newpage
\setcounter{page}{1}

\tableofcontents\newpage


\section{Introduction}

A fundamental observation of nature is that electroweak interactions act on chiral fermions and
treat left- and right-handed fermions differently.
Setting up a consistent regularisation and renormalisation of the relevant chiral gauge theories, however, 
proves to be difficult. 
In particular, dimensional regularisation (DReg) of chiral gauge theories inevitably leads to 
the so-called $\gamma_5$-problem emerging from the challenge of accommodating the manifestly 
4-dimensional nature of $\gamma_{5}$ in $D$ dimensions, as already discussed early on e.g.\ in Refs.\ 
\cite{tHooft:1972tcz,Bollini:1972ui,Cicuta:1972jf,Akyeampong:1973xi,Akyeampong:1973vk,Akyeampong:1973vj,Chanowitz:1979zu,Trueman:1995ca,Jegerlehner:2000dz}.
Still, DReg is the most commonly used scheme for practical calculations, because
it allows efficient practical computations and satisfies causality, Lorentz invariance and unitarity; 
for a review of variants of DReg and alternatives we refer to Ref.\ \cite{Gnendiger:2017pys}.

A rigorous and consistent way to embed $\gamma_5$ into the framework of DReg is the 
Breitenlohner-Maison/'t Hooft-Veltman (BMHV) scheme
\cite{tHooft:1972tcz,Breitenlohner:1975hg,Breitenlohner:1976te,Breitenlohner:1977hr},
which abandons the anticommutativity of $\gamma_5$.
Employing the BMHV scheme, however, violates gauge invariance in intermediate steps of the
regularisation and renormalisation procedure due to the modified algebra.
This is a spurious breaking, which can and needs to be
restored using symmetry-restoring counterterms guaranteed to exist (in anomaly free theories) by the methods 
of algebraic renormalisation \cite{Piguet:1995er}.
However, as these counterterms are gauge non-invariant, including evanescent operators,
they cannot be generated via a multiplicative renormalisation transformation, making the 
traditional text book approach to renormalisation insufficient and leading to a more general
but also more complicated counterterm structure.

Here, we focus on the mathematically rigorous
BMHV scheme in chiral gauge theories and the required counterterm structure. 
Previous publications \cite{Belusca-Maito:2020ala,Belusca-Maito:2021lnk,Belusca-Maito:2022wem} 
covered non-abelian chiral gauge theories with fermions and scalars at the one-loop level and 
an abelian chiral gauge theory at the two-loop level and obtained all required counterterms and 
renormalisation-group beta-functions. 
The theoretical basis and the methodology of these papers is reviewed in detail 
in Ref.\ \cite{Belusca-Maito:2023wah}. 
Related analyses of the BMHV scheme covered Yang-Mills theories without scalars at the one-loop 
level \cite{Martin:1999cc}, also using the background-field gauge 
\cite{Cornella:2022hkc}, the abelian Higgs model at the one-loop level \cite{Sanchez-Ruiz:2002pcf}, the two-loop computation of beta-functions in non-gauge theories
\cite{Schubert:1988ke}, and recent applications to effective theories
\cite{Naterop:2023dek,DiNoi:2023ygk}.

In the present paper we present the first application of the BMHV scheme to a chiral gauge theory 
at the three-loop level. 
The paper is a direct continuation of the previous publication \cite{Belusca-Maito:2021lnk} 
and studies an abelian chiral gauge theory, which serves as a toy model for the investigation 
of theoretical concepts. We obtain a consistently renormalised finite theory with restored BRST
invariance at the three-loop level. 
All required counterterms and Green functions are explicitly provided.  
Ultimately, such a renormalisation procedure will be needed for high-precision 
calculations of e.g.\ electroweak observables in the BMHV scheme.

We briefly comment on interesting alternative schemes for $\gamma_5$. 
An important alternative scheme 
for the treatment of $\gamma_5$ is "Kreimer's scheme"
\cite{Kreimer:1989ke,Korner:1991sx,Kreimer:1993bh}, promising a better behaviour with respect to 
gauge invariance by abandoning the cyclicity of the trace.
However, the multi-loop properties of this scheme are not fully under control. 
Refs.\ \cite{Bednyakov:2015ooa,Zoller:2015tha,Poole:2019txl,Davies:2019onf,Davies:2021mnc,Herren:2021vdk}
showed that there are ambiguities in some of
the $\beta$-function coefficients which had to be fixed by external arguments using Weyl
consistency conditions
\cite{Osborn:1989td,Jack:1990eb,Osborn:1991gm,Jack:2013sha,Poole:2019txl,Poole:2019kcm} 
and Ref.\ \cite{Chen:2023ulo} showed
that, in the context of higher order QCD corrections with an external flavour-singlet axial-current, 
the ABJ equation \cite{Adler:1969gk,Bell:1969ts} and the Adler-Bardeen theorem \cite{Adler:1969er} 
do not automatically hold
in the bare form when treating $\gamma_5$ in Kreimer's scheme, but in fact and contrary to expectations,
additional counterterms were needed making the traditional multiplicative renormalisation
insufficient in this scheme as well. Finally, we mention Ref.\ \cite{Bruque:2018bmy} which showed
that even if DReg is entirely abandoned and purely 4-dimensional regularisation schemes are 
considered, an analog of the $\gamma_5$ problem exists in a very broad class of potential 
regularisation schemes.

In Sec.\ \ref{Sec:ModelAndMethodology} of the present paper, we introduce the considered 
abelian model and briefly sketch the methodology behind the computation of the 
aforementioned symmetry-restoring counterterms via special Feynman 
diagrams with an insertion of the $\widehat{\Delta}$-operator using the regularised 
quantum action principle of DReg.
In order to extract UV-divergences at the multi-loop level an infrared rearrangement via the 
so called all massive tadpoles method \cite{Misiak:1994zw,Chetyrkin:1997fm} is utilised, 
where all occurring Feynman diagrams are mapped to fully massive single-scale vacuum bubbles.
We explain this method in Sec.\ \ref{Sec:AMT}.
Finally, we present and discuss the new three-loop level results in Sec.\ \ref{Sec:3-Loop-Ren-Results},
before we conclude in the last section \ref{Sec:Conclusion},
showing that the counterterm structure in the BMHV scheme may still be written in a 
rather compact form, suitable for computer implementations, even at high loop levels.
In the appendix we provide explicit results for the three-loop coefficients, introduced in 
Sec.\ \ref{Sec:3-Loop-Ren-Results}, and further display the one- and two-loop results for completeness,
already published in Ref.\ \cite{Belusca-Maito:2021lnk}, this time, however, in full $R_{\xi}$-gauge.


\section{Abelian chiral Gauge Theory and Dimensional Regularisation with non-anticommuting $\gamma_5$}\label{Sec:ModelAndMethodology}

In this section we briefly introduce the BMHV algebra with non-anticommuting $\gamma_5$, 
the considered abelian chiral gauge theory and its definition in $D$ dimensions
and eventually the methodology of the symmetry restoration procedure. 
The model is the same as the one discussed at the two-loop level in Ref.\ 
\cite{Belusca-Maito:2021lnk}. For a more detailed review of the basic methodology including 
applications to the abelian chiral gauge theory we also refer to
Ref.\ \cite{Belusca-Maito:2023wah}, particularly to Sec.\ 
$3.3$ and $7.2$, as well as Sec.\ $4$ and $6.3$ regarding the definition of the regularisation, 
and the theory of symmetry restoration.

\subsection{Breitenlohner-Maison/’t Hooft-Veltman Algebra}\label{Sec:BMHV-Algebra}

As already mentioned in the introduction, 
anticommutativity of $\gamma_5$ is abandoned in the BMHV scheme in
order to obtain a consistent dimensional regularisation of a chiral gauge theory, which leads to
modified algebraic relations, the so-called BMHV algebra
\begin{equation}\label{Eq:BMHVAlgebra}
    \begin{aligned}
        \{\gamma_5, \overline{\gamma}^{\mu}\} &= 0,
        &\hspace{1cm}
        \{\gamma_5, \gamma^{\mu}\} &= \{\gamma_5, \hat{\gamma}^{\mu}\} = 2 \, \gamma_5 \, \hat{\gamma}^{\mu},\\
        [\gamma_5, \hat{\gamma}^{\mu}] &= 0,
        &\hspace{1cm}
        [\gamma_5, \gamma^{\mu}] &= [\gamma_5, \overline{\gamma}^{\mu}] = 2 \, \gamma_5 \, \overline{\gamma}^{\mu},
    \end{aligned}
\end{equation}
where the $D$-dimensional space as well as all Lorentz covariants are decomposed into a $4$- and
a $(-2\epsilon)$-dimensional component as
\begin{equation}
    \begin{aligned}
        &\mathbb{M}=\mathbb{M}_4\oplus\mathbb{M}_{-2\epsilon},
        &\hspace{1cm}
        &\eta_{\mu\nu} = \overline{\eta}_{\mu\nu} + \hat{\eta}_{\mu\nu},
        &\hspace{1cm}
        &X^{\mu} = \overline{X}^{\mu} + \hat{X}^{\mu},
    \end{aligned}
\end{equation}
with overbars and hats denoting $4$-dimensional and $(-2\epsilon)$-dimensional 
components, respectively.
These modified algebraic relations (\ref{Eq:BMHVAlgebra}) are the root of the spurious
BRST symmetry-breaking in intermediate steps. As an illustration we consider a generic kinetic term 
of a fermion field $\psi$. Splitting the fermion field into its left-handed and right-handed parts as 
$\psi=\psi_L+\psi_R=\mathbb{P}_{\mathrm{L}} \psi+\mathbb{P}_{\mathrm{R}} \psi$ with the projectors 
$\mathbb{P}_{\mathrm{L,R}} =(1\mp\gamma_5)/2$, the $D$-dimensional kinetic term decomposes as
\begin{align}\label{Eq:illustratekineticterm}
    \overline{\psi} {\slashed{\partial}} \psi= \overline{\psi_L} \overline{\slashed{\partial}} \psi_L+\overline{\psi_R} \overline{\slashed{\partial}} \psi_R
    +\overline{\psi_L} \widehat{\slashed{\partial}} \psi_R+\overline{\psi_R} \widehat{\slashed{\partial}} \psi_L
\end{align}
into four terms. The first two involve a 4-dimensional derivative and do not mix chiralities. 
The last two, however, involve a $(-2\epsilon)$-dimensional derivative and mix chiralities. 
In a chiral gauge theory such terms violate gauge invariance.

\subsection{Abelian chiral Gauge Theory and its Extension to $D$ Dimensions} \label{Sec:ToDandBeyond}

The considered abelian chiral gauge theory is the same as the one considered in 
Refs.\ \cite{Belusca-Maito:2021lnk,Belusca-Maito:2023wah} and involves an abelian 
gauge field $A_{\mu}$ which interacts only with the right-handed fermions $\psi_R$. 
Details on the construction of the 4-dimensional formulation, symmetry identities 
and the extension to $D$ dimensions can be found in the literature.
Its gauge covariant derivative is written as\footnote{
Note the change of the sign convention in the covariant derivative compared
to Ref.\ \cite{Belusca-Maito:2021lnk}, affecting also some of the results for the 
counterterms. Here, we use the same convention as in Ref.\ \cite{Belusca-Maito:2023wah}.}
\begin{equation}
    \begin{aligned}
        D_{ij}^\mu = \partial^\mu \delta_{ij} + i e A^\mu {\mathcal{Y}_R}_{ij}
    \end{aligned}
\end{equation}
with hypercharge matrix ${\mathcal{Y}_R}_{ij}$ which must satisfy the anomaly
cancellation condition
\begin{equation}\label{Eq:AnomalyCancellationCondition}
    \begin{aligned}
        \text{Tr}\big(\mathcal{Y}_R^3\big) = 0,
    \end{aligned}
\end{equation}
to guarantee a consistent theory.
The field strength tensor takes the usual abelian form
\begin{equation}
    \begin{aligned}
        F_{\mu\nu} = \partial_{\mu} A_{\nu} - \partial_{\nu} A_{\mu}.
    \end{aligned}
\end{equation}
The BRST transformations of the gauge and matter fields, the Faddeev-Popov ghosts and 
antighosts $c,\bar{c}$ and the Nakanishi-Lautrup field $B$ are given by
\begin{equation}\label{Eq:BRST-Trafos}
    \begin{aligned}
        s A_{\mu}(x) &= \partial_{\mu} c(x),\\
        s\psi_{i}(x) &= s{\psi_R}_i(x) = - i e c(x) {\mathcal{Y}_R}_{ij} {\psi_R}_j(x),\\
        s\overline{\psi}_i(x) &= s\overline{\psi_R}_i(x) = - i e \overline{\psi_R}_j(x) c(x) {\mathcal{Y}_R}_{ji} = i e c(x) \overline{\psi_R}_j(x) {\mathcal{Y}_R}_{ji},\\
        s c(x) &= 0,\\
        s\bar{c}(x) &= B(x) = - \frac{1}{\xi} \partial^{\mu}A_{\mu}(x),\\
        s B(x) &= 0,
    \end{aligned}
\end{equation}
with $s$ being the BRST operator.
We stress that only the right-handed component of the fermions admits non-trivial 
BRST transformations.

All previous equations can readily be interpreted in $D$ dimensions.
In defining the $D$-dimensional regularised action, however, one faces two major challenges.
First, the fermionic kinetic term must be fully $D$-dimensional as in 
Eq.\ (\ref{Eq:illustratekineticterm}) 
in order to generate propagators with $D$-dimensional denominators and correctly regularised loop 
Feynman diagrams. Thus, the kinetic term necessarily mixes fermions of different chiralities. 
In the present case only the right-handed fermion has non-vanishing gauge and BRST
transformations and couples to the gauge boson via its hypercharge, while the left-handed fermion
is sterile, i.e.\ has no interactions and vanishing BRST transformation. 
This left-handed fermion $\psi_L$ is purely fictitious and only introduced 
for the $D$-dimensional formulation and it only
appears in the kinetic term. 
Second, the extension to $D$ dimensions is not
unique due to the fact that the right-handed chiral current 
$\overline{\psi_R}_i \gamma^{\mu} {\psi_R}_j$ admits many inequivalent but equally correct
extensions. 
Here, we follow Ref.\ \cite{Belusca-Maito:2021lnk} and use the most symmetric option 
$\overline{\psi} \mathbb{P}_{\mathrm{L}} \gamma^{\mu} \mathbb{P}_{\mathrm{R}} \psi
= \overline{\psi} \mathbb{P}_{\mathrm{L}} \overline{\gamma}^{\mu} \mathbb{P}_{\mathrm{R}} \psi
= \overline{\psi_{R}} \overline{\gamma}^{\mu} \psi_{R}$,
as it is the most natural and symmetric choice. This choice was also made in most BMHV applications 
to chiral gauge theories in the literature 
\cite{Belusca-Maito:2020ala,Belusca-Maito:2021lnk,Belusca-Maito:2023wah,Martin:1999cc,
Sanchez-Ruiz:2002pcf,Cornella:2022hkc,DiNoi:2023ygk} and 
likely leads to the simplest expressions particularly in applications to the 
electroweak SM \cite{Jegerlehner:2000dz}.

The $D$-dimensional tree-level action may then be written as
\begin{equation}\label{Eq:D-dimAction}
    \begin{aligned}
        S_{0} &= \Dintx \, \bigg( i \overline{\psi}_i \slashed{\partial} {\psi}_i - e {\mathcal{Y}_R}_{ij} \overline{\psi_R}_i \slashed{A} {\psi_R}_j - \frac{1}{4} F^{\mu\nu} F_{\mu\nu}\\
        &\hspace{1.75cm} - \frac{1}{2 \xi} (\partial_\mu A^\mu)^2 - \bar{c}\partial^2c + \rho^\mu s{A_\mu} + \bar{R}^i s{{\psi_R}_i} + R^i s{\overline{\psi_R}_i} \bigg).
    \end{aligned}
\end{equation}
In this action, all derivatives are continued in the obvious way to $D$ dimensions. 
However, we emphasise that in the chosen fermion-gauge boson interaction only the purely 4-dimensional 
gamma matrices appear due to the projection operators. The BRST transformations are coupled to 
external sources $\rho^\mu$, $\bar{R}^i$ and $R^i$.

Preparing for the renormalisation and higher orders, the BRST transformations are replaced by a 
Slavnov-Taylor operator $\mathcal{S}_D$, whose explicit definition can be found in Refs.\ 
\cite{Belusca-Maito:2020ala,Belusca-Maito:2021lnk,Belusca-Maito:2023wah}. 
Acting with this $D$-dimensional Slavnov-Taylor operator
on the $D$-dimensional tree-level action to check for BRST invariance, we obtain
\begin{equation}\label{Eq:SDbreaking}
    \begin{aligned}
        \mathcal{S}_{D}(S_{0}) = \mathcal{S}_{D}(S_{0,\text{evan}}) = \widehat{\Delta},
    \end{aligned}
\end{equation}
where in the second equality the purely evanescent kinetic term has been introduced as
\begin{equation} \label{Eq:S_0,evan}
    \begin{aligned}
        S_{0,\text{evan}} = \Dintx \, i \overline{\psi}_i \widehat{\slashed{\partial}} \psi_i
    \end{aligned}
\end{equation}
and where the breaking operator $\widehat{\Delta}$ has the explicit form
\begin{equation} \label{Eq:Delta_Hat}
    \begin{aligned}
        \widehat{\Delta} = - \Dintx \, e \,  {\mathcal{Y}_R}_{ij} \, c \, 
                    \bigg\{
                    \overline{\psi}_i \bigg(
                    \overset{\leftarrow}{\widehat{\slashed{\partial}}} \mathbb{P}_{\mathrm{R}} 
                    + \overset{\rightarrow}{\widehat{\slashed{\partial}}} \mathbb{P}_{\mathrm{L}}
                    \bigg) \psi_j 
                    \bigg\}
                = \Dintx \, \widehat{\Delta}(x).
    \end{aligned}
\end{equation}
The non-vanishing result for the quantity $\widehat{\Delta}$ corresponds to the 
announced breaking of BRST invariance by the BMHV scheme. The breaking happens already at 
the level of the tree-level action. The second equation in (\ref{Eq:SDbreaking})  shows
that the breaking is caused only by the evanescent part of the kinetic term given in 
Eq.\ (\ref{Eq:S_0,evan}). This term
is required in order to formulate a
$D$-dimensional fermion propagator, but as already illustrated in 
Eq.\ (\ref{Eq:illustratekineticterm}) it mixes left- and 
right-handed fields with different gauge transformation properties, which is the technical 
reason for the breaking of BRST invariance in the BMHV scheme.

The BRST breaking can also be viewed as a composite operator. From the explicit form 
in Eq.\ (\ref{Eq:Delta_Hat}) we can derive
a Feynman rule for insertions of the $\widehat{\Delta}$-operator, which 
takes the following form
\begin{equation}
    \begin{tabular}{rl}
        \raisebox{-42pt}{\includegraphics[scale=0.65]{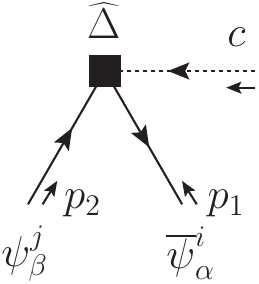}}&
        $\begin{aligned}
            &= - e \, {\mathcal{Y}_R}_{ij} \left(\widehat{\slashed{p}}_1 \mathbb{P}_{\mathrm{R}} + \widehat{\slashed{p}}_2 \mathbb{P}_{\mathrm{L}} \right)_{\alpha\beta}.
        \end{aligned}$
    \end{tabular}
\end{equation}

To conclude this subsection, we note that, in contrast to the non-abelian case, 
in abelian gauge theories none of the BRST
transformations and none of the terms in the second line of the action (\ref{Eq:D-dimAction}) 
above obtain quantum corrections, i.e.\ they all do not renormalise. 
This is due to the fact that all these terms are bilinear in the quantum fields. 
This non-renormalisation can be formulated as a local Ward identity or as an antighost equation.
In the following we can assume the corresponding relations to be valid without further discussion 
as long as it is made sure not to violate the relations by an inappropriate choice of 
symmetry-restoring counterterms. Hence, we
do not have to consider
Green functions with external sources, which would be necessary in the non-abelian case.
For more details regarding these issues we refer the reader to Sec.\ 2.6 of Ref.\ 
\cite{Belusca-Maito:2023wah} and references therein.

\subsection{Procedure of Symmetry Restoration} \label{Sec:SymmetryRestoration}

At the quantum level, the theory is regularised and renormalised using DReg and the BMHV scheme. 
In this procedure, a counterterm action $S_\mathrm{ct}$ is added to the tree-level action $S_0$ which 
cancels UV-divergences and spurious breakings of BRST invariance. The renormalised theory is described 
by the effective quantum action in $D$ dimensions  $\Gamma_\mathrm{DRen}$, which is also the 
generating functional of one-particle irreducible (1PI) Green functions. The final 4-dimensional 
renormalised effective action is then obtained as 
$\Gamma=\mathop{\text{LIM}}_{D \, \to \, 4}\Gamma_\mathrm{DRen}$,
where the operation $\mathop{\text{LIM}}_{D \, \to \, 4}$ means taking the $D=4$ limit and 
neglecting algebraic terms which are evanescent, i.e.\ which vanish in 4 dimensions.
The ultimate symmetry requirement is the Slavnov-Taylor identity, which needs to be satisfied by
our theory after renormalisation and in $4$ dimensions, i.e.\
\begin{equation}\label{Eq:UltimateSymmetryRequirement}
    \begin{aligned}
        \mathop{\text{LIM}}_{D \, \to \, 4} \, (\mathcal{S}_D(\Gamma_\mathrm{DRen})) = 0.
    \end{aligned}
\end{equation}

The $\widehat{\Delta}$-operator becomes of particular importance for the symmetry restoration.
In the present case where the classical symmetry is broken by the regularisation, we employ
the regularised quantum action principle of DReg (see Ref.\ \cite{Breitenlohner:1977hr} and 
also the review \cite{Belusca-Maito:2023wah})
\begin{equation}\label{Eq:QAPofDReg}
    \begin{aligned}
        \mathcal{S}_D(\Gamma_\mathrm{DRen})=(\widehat{\Delta}+\Delta_{\mathrm{ct}})\cdot\Gamma_\mathrm{DRen},
    \end{aligned}
\end{equation}
in order to rewrite a possible symmetry-breaking as a composite operator insertion into the
effective quantum action.
The operator $\widehat{\Delta}$ has been defined above, and the operator $\Delta_{\mathrm{ct}}$ is 
obtained similarly by the violation of the Slavnov-Taylor identity of the action including counterterms as
\begin{equation}
    \begin{aligned}
\widehat{\Delta} + \Delta_{\mathrm{ct}} = \mathcal{S}_D(S_{0} + S_{\mathrm{ct}}).
    \end{aligned}
\end{equation}

Practically, we can plug Eq.\ (\ref{Eq:QAPofDReg}) into Eq.\ (\ref{Eq:UltimateSymmetryRequirement})
to obtain the following perturbative requirement from the Slavnov-Taylor identity
\begin{equation}\label{Eq:PerturbativeRequirementAndStartingPoint}
    \begin{aligned}
        \mathop{\text{LIM}}_{D \, \to \, 4} \, \bigg(\widehat{\Delta}\cdot\Gamma_\mathrm{DRen}^n+\sum_{k=1}^{n-1}\Delta^k_\mathrm{ct}\cdot\Gamma^{n-k}_\mathrm{DRen}+\Delta^n_\mathrm{ct}\bigg)=0, \hspace{0.75cm} \forall \, n \geq 1,
    \end{aligned}
\end{equation}
with $n$ being the loop order of the respective quantities.
This equation can be used as the starting point of the iterative symmetry restoration procedure. 
Supposing that the theory has been renormalised up to some loop order $n-1$, the counterterm action
$S_\mathrm{ct}^{n-1}$ and the corresponding breaking operator $\Delta_\mathrm{ct}^{n-1}$ are known 
up to order $n-1$.
The first two terms in Eq.\ (\ref{Eq:PerturbativeRequirementAndStartingPoint}) can then be 
unambiguously computed at the next order $n$. At this order $n$, the counterterm action 
$S_\mathrm{ct}^n$ then needs to be determined such that the third term cancels the first two and  
Eq.\ (\ref{Eq:PerturbativeRequirementAndStartingPoint}) is fulfilled.

The symmetry restoration thus requires the computation of subrenormalised 1PI Green functions with 
one insertion of the operator $\Delta=\widehat{\Delta}+\Delta_\mathrm{ct}$, whose lowest-order part 
is evanescent.
The main advantage of this method is its efficiency, as only power-counting divergent
Green functions need to be considered and only their UV-divergent part needs to be
calculated, which is a crucial feature at higher loop orders. The method has been applied at lower orders 
in Refs.\ \cite{Belusca-Maito:2020ala,Belusca-Maito:2021lnk,Martin:1999cc,Cornella:2022hkc} 
(for a three-loop application in the context of supersymmetry see Ref.\ \cite{Stockinger:2018oxe}), 
and a detailed comparison with alternative methods have been given in 
Refs.\ \cite{Belusca-Maito:2020ala,Belusca-Maito:2023wah}.


\section{Extracting UV-Divergences at the Multi-Loop Level}\label{Sec:AMT}

We need to calculate not only the symmetric and non-symmetric (i.e.\ symmetry-breaking)
UV-divergences, but also the finite symmetry-breaking contributions.
The latter are local contributions obtained from the UV-divergences of 1PI
Green functions with the insertion of an evanescent operator, as already mentioned at the
end of the previous section. Thus, utilising the quantum action principle, only 
the UV-divergent part of power-counting divergent 1PI Green functions needs to be computed.
For the present work we did this up to the three-loop level; the results will be displayed below in Sec.\ 
\ref{Sec:3-Loop-Ren-Results}.

The computations are mainly performed in \texttt{Mathematica} \cite{mathematica}, but partly also in
\texttt{C++} for the integral reduction. In particular, the \texttt{Mathematica} package 
\texttt{FeynArts} \cite{Hahn:2000kx}
has been used to generate all Feynman diagrams, including diagrams with insertions of the operator 
$\Delta$ and/or the $(\le2)$-loop counterterms given in the Appendix. Most symbolic manipulations,
especially those related to the Dirac algebra, have been performed with the help of
\texttt{FeynCalc} \cite{Mertig:1990an, Shtabovenko:2016sxi, Shtabovenko:2020gxv, Shtabovenko:2021hjx}.
Further, the package \texttt{FeynHelpers} \cite{Shtabovenko:2016whf} has been used to
interface the \texttt{Mathematica} setup with the \texttt{C++} version of the
software \texttt{FIRE} \cite{Smirnov:2019qkx}, which uses integration by parts (IBP) identities to 
reduce all Feynman integrals
to master integrals.

Note that renormalising chiral gauge theories with 
non-anticommuting $\gamma_5$ in the BMHV scheme means that we cannot use 
Ward or Slavnov-Taylor identities to circumvent 
the calculation of multi-leg 1PI Green functions as it is 
usually done 
(see e.g.\ \cite{Davies:2019onf,Davies:2021mnc,Herren:2021vdk,Herzog:2017ohr,Luthe:2017ttg}), 
because gauge invariance is broken
in intermediate steps by the regularisation.
In other words, we also have to calculate all divergent three- and four-point Green functions. 

Hence, we aim for a method to reduce all integrals to fully massive single-scale ones, 
not only to drastically reduce the computational complexity, or to even be able to find solutions
for the master integrals at all, but also to avoid possible IR-divergences. 
Noting that counterterms are local polynomials in external momenta and 
(for mass-independent schemes) internal masses,
we can extract the UV-divergences utilizing an infrared rearrangement to achieve this task.
In particular, we are using the all massive tadpoles method, first introduced in Refs.\
\cite{Misiak:1994zw,Chetyrkin:1997fm}, where the infrared rearrangement is realised via the
following exact decomposition
\begin{equation} \label{Eq:StandardTadpoleExpansion}
    \begin{aligned}
        \frac{1}{(k+p)^2} = \frac{1}{k^2 - M^2} - \frac{p^2 + 2 \, k \cdot p + M^2}{k^2 - M^2} \, \frac{1}{(k+p)^2},
    \end{aligned}
\end{equation}
where $k$ is a loop momentum or any linear combination of loop momenta.
This decomposition can be applied recursively up to a sufficient order, 
given by the corresponding degree of divergence.
Power-counting finite terms may then be dropped, which does not affect the UV-divergences 
after proper subtraction of subdivergences.
In this way, having introduced the auxiliary mass scale $M^2$, which is the only scale remaining
in the denominators, all occurring Feynman diagrams
are mapped to fully massive single-scale vacuum bubbles.

However, as discussed in Ref.\ \cite{Lang:2020nnl}, we also found that there are some subtleties
w.r.t.\ the tadpole expansion in Eq.\ (\ref{Eq:StandardTadpoleExpansion}) when applied to two and 
higher loop orders due to subdivergences. 
The application of this tadpole expansion requires a one-to-one correspondence
between the expansions of the integrals of the genuine $L$-loop diagrams and the integrals of 
their corresponding counterterm-inserted diagrams with lower loop level.
The reason for this is that separately, they are not momentum routing independent anymore
after truncating the tadpole expansion (\ref{Eq:StandardTadpoleExpansion}); only their combination
is momentum routing independent.

As this is not convenient for computer implementations, we decided to use an
improved tadpole expansion, as already implied in Refs.\ \cite{Misiak:1994zw,Chetyrkin:1997fm}
and explained in Ref.\ \cite{Lang:2020nnl}.
Here, the auxiliary mass scale $M^2$ is introduced in every propagator
and subsequently a Taylor-expansion in external momenta 
(and in internal/physical masses if they were present) 
is performed. Exemplarily, for one massless propagator, we obtain
\begin{equation}
    \begin{aligned}
        \frac{1}{(k+p)^2} \longrightarrow \frac{1}{(k+p)^2 - M^2} = \frac{1}{k^2 - M^2} - \frac{p^2 + 2 \, k \cdot p}{(k^2 - M^2)^2}  + \frac{(p^2 + 2 \, k \cdot p)^2}{(k^2 - M^2)^3} + \ldots,
    \end{aligned}
\end{equation}
where it can be seen that the same result as with the exact decomposition is obtained 
when neglecting numerator terms $\propto M^2$ in Eq.\ (\ref{Eq:StandardTadpoleExpansion}).
However, neglecting such numerator terms $\propto M^2$ needs to be compensated, in particular
at the multi-loop level with occurring subdivergences.
This is done by constructing and including all possible auxiliary counterterms which 
are $\propto M^2$
at a given order. In the present case such auxiliary counterterms can correspond to 
mass terms of the 4-dimensional gauge field or the evanescent gauge field $\hat{A}^\mu$, 
or to counterterms appearing in the renormalisation of the 
insertion operator $\widehat{\Delta}$, see Sec.\
\ref{Sec:ToDandBeyond} and Sec.\ \ref{Sec:3-Loop-BRST-Breaking}.
Both the auxiliary mass $M^2$ and the auxiliary mass counterterms are only present 
at the level of the
Feynman integral evaluation and are not part of the theory; hence, they may be viewed as a
mathematical trick.
In particular, an auxiliary gauge boson mass counterterm, 
cf.\ Refs.\  \cite{Misiak:1994zw,Chetyrkin:1997fm},
does not represent a problem.

After all Feynman integrals have been mapped to these fully massive single-scale vacuum bubbles,
they are reduced to a finite set of master integrals via IBP-relations using
\texttt{FIRE}. The required
solutions for the two two-loop and the five three-loop master integrals have been taken from Refs.\
\cite{Schroder:2005va, Martin:2016bgz}.


\section{Three-Loop Renormalisation: Evaluation of the singular and finite Counterterm Action}\label{Sec:3-Loop-Ren-Results}

In this section we present the complete three-loop renormalisation of the considered
abelian chiral gauge theory, regularised in DReg and using the BMHV scheme.
We first compute the results of all required 1PI Green functions, including Green functions 
with $\Delta$ insertions describing the breaking of the Slavnov-Taylor identity. 
From the UV-divergences of the Green functions we derive the corresponding UV-divergent
counterterm action; and from the breaking of the Slavnov-Taylor identity we derive the 
symmetry-restoring counterterms.

In order to highlight the structure of the three-loop results, they are presented in terms 
of abbreviations which are defined in App.\ \ref{App:3-Loop-Coeffs}. 
The corresponding one- and two-loop results are listed for completeness 
in App.\ \ref{App:1-Loop-Results} and App.\ \ref{App:2-Loop-Results}, respectively. 
For those results we go beyond the literature and provide them for general gauge parameter $\xi$.
The three-loop results are provided in Feynman gauge $\xi=1$.

All results have been obtained using the computational setup described in Sec.\ \ref{Sec:AMT},
which has successfully been tested using standard vector-like quantum electrodynamics
by performing a complete three-loop renormalisation.
Further, the one- and two-loop results for the chiral model considered here, published in Ref.\
\cite{Belusca-Maito:2021lnk}, have successfully been reproduced.
For the three-loop results, the UV-divergent BRST breaking contributions can be 
obtained on the one hand from standard 1PI Green functions, see Sec.\ 
\ref{Sec:3-Loop-Div-GreenFuncs}, and on the other hand from the 
$\Delta$-inserted 1PI Green functions, 
cf.\ Sec.\ \ref{Sec:3-Loop-BRST-Breaking}. The results agree, serving as a 
strong consistency check.
Moreover, all obtained counterterms, including the finite symmetry-restoring ones, 
are local polynomials in the external momenta, as expected. The observed cancellation
of logarithmic terms depends critically on all details of the implementation of
lower-order counterterms such as the dimensionality 
(i.e.\ either $D$-, $4$- or $(-2\epsilon)$-dimensional) of all appearing Lorentz structures, 
and thus gives us further
confidence in the correctness of the results.

\subsection{Divergent Three-Loop Green Functions}\label{Sec:3-Loop-Div-GreenFuncs}

We begin with all standard, i.e.\ non-operator inserted, 1PI Green functions,
which can possibly lead to UV-divergences. 
From these we will later derive all singular counterterms to ultimately render the theory finite. 
The complete list of relevant Green functions is as follows.

\vspace{0.2cm}\noindent\textbf{(i) Gauge Boson Self Energy:}\vspace{0.1cm}

\noindent The divergent part of the three-loop gauge boson self energy (after subrenormalization using one- and two-loop counterterms from the literature and reproduced in the Appendix) is given by
\begin{equation}
    \begin{aligned}
        i \widetilde{\Gamma}_{AA}^{\nu\mu}(p) \big|_{\text{div}}^{3} = 
        &- \frac{i}{(16 \pi^2)^3} \, e^6 \, 
        \bigg[ \mathcal{B}_{AA}^{3, \text{inv}} \, \frac{1}{\epsilon^2} + \mathcal{A}_{AA}^{3, \text{inv}} \, \frac{1}{\epsilon} \bigg]
        \Big(\overline{p}^{\mu} \overline{p}^{\nu} - \overline{p}^2 \overline{\eta}^{\mu\nu}\Big)\\
        &- \frac{i}{(16 \pi^2)^3} \, e^6 \,
        \bigg[ \widehat{\mathcal{C}}_{AA}^{3, \text{break}} \, \frac{1}{\epsilon^3} + \widehat{\mathcal{B}}_{AA}^{3, \text{break}} \, \frac{1}{\epsilon^2} + \widehat{\mathcal{A}}_{AA}^{3, \text{break}} \, \frac{1}{\epsilon} \bigg] \, \widehat{p}^2 \, \overline{\eta}^{\mu\nu}\\
        &+ \frac{i}{(16 \pi^2)^3} \, e^6 \, \overline{\mathcal{A}}_{AA}^{3, \text{break}} \, \frac{1}{\epsilon} \, \overline{p}^2 \, \overline{\eta}^{\mu\nu},
    \end{aligned}
\end{equation}
with three-loop coefficients provided in Eqs.\ (\ref{Eq:BphotonInv}) 
to (\ref{Eq:AphotonBreakNonEvan}) in appendix \ref{App:3-Loop-Coeffs}.
The first line is the expected transverse part, here written with purely 4-dimensional covariants.
The second line breaks transversality by an evanescent operator, which already appears at the 
one- and two-loop level, see Eqs.\ (\ref{Eq:Ssct_1-Loop}, \ref{Eq:Ssct_2-Loop}) and 
Ref.\ \cite{Belusca-Maito:2021lnk}. In contrast, the third line contains a non-evanescent but 
UV-divergent BRST breaking contribution, which for the first time appears at the three-loop level.

\vspace{0.2cm}\noindent\textbf{(ii) Fermion Self Energy:}\vspace{0.1cm}

\noindent The UV-divergences of the fermion self energy at three loop order are provided by
\begin{equation}\label{Eq:FermionSelfEnergy_3-Loop}
    \begin{aligned}
        i \widetilde{\Gamma}_{\psi\overline{\psi}}^{ji}(p) \big|_{\text{div}}^{3} &= \frac{i}{(16 \pi^2)^3} \, e^6 \, 
        \bigg\{ {\mathcal{C}}_{\psi\overline{\psi}}^{3, \, ji} \, \frac{1}{\epsilon^3} + {\mathcal{B}}_{\psi\overline{\psi}}^{3, \, ji} \, \frac{1}{\epsilon^2} + {\mathcal{A}}_{\psi\overline{\psi}}^{3, \, ji} \, \frac{1}{\epsilon} \bigg\} \, \overline{\slashed{p}} \, \mathbb{P}_{\mathrm{R}}\\
        &= \frac{i}{(16 \pi^2)^3} \, e^6 \, 
        \bigg\{ {\mathcal{C}}_{\psi\overline{\psi}, \, ji}^{3, \, \text{inv}} \, \frac{1}{\epsilon^3} + \Big( {\mathcal{B}}_{\psi\overline{\psi}, \, ji}^{3, \, \text{inv}} + {\mathcal{B}}_{\psi\overline{\psi}, \, ji}^{3, \, \text{break}} \Big) \, \frac{1}{\epsilon^2}\\
        &\hspace{4.09cm} + \Big( {\mathcal{A}}_{\psi\overline{\psi}, \, ji}^{3, \, \text{inv}} + {\mathcal{A}}_{\psi\overline{\psi}, \, ji}^{3, \, \text{break}} \Big) \, \frac{1}{\epsilon} \bigg\} \, \overline{\slashed{p}} \, \mathbb{P}_{\mathrm{R}},
    \end{aligned}
\end{equation}
with three-loop coefficients to be found in Eqs.\ (\ref{Eq:CfermionInv}) 
to (\ref{Eq:AfermionBreak}) in appendix \ref{App:3-Loop-Coeffs}.
In the second and third line, the result has been split into invariant contributions 
and contributions which break BRST invariance. 
The split is related to the fermion-gauge boson three-point function discussed next. 
The breaking terms in Eq.\ (\ref{Eq:FermionSelfEnergy_3-Loop}) correspond to the violation of 
the well-known Ward identity relating the fermion self energy and the fermion-gauge boson
interaction in an abelian gauge theory, and we follow the convention used already at 
lower orders in Ref.\ \cite{Belusca-Maito:2021lnk} to attribute the entire breaking of 
this Ward identity to the fermion self energy.

Whereas in the one-loop case there is no such UV divergent breaking contribution 
in the fermion self energy, 
cf.\ Eq.\ (\ref{Eq:Ssct_1-Loop}),
and in the two-loop case there is only a breaking contribution coming from the 
simple $\epsilon$-pole, cf.\ Eq.\ (\ref{Eq:Ssct_2-Loop}), 
here in the three-loop case there is
also a symmetry-violating contribution from the second order $\epsilon$-pole. 
I.e.\ ${\mathcal{B}}_{\psi\overline{\psi}, \, ji}^{3, \, \text{break}}$ starts being 
non-zero at the three-loop level.
Again, the complete BRST breaking contribution from the fermion self energy is, 
as in the two-loop case, purely non-evanescent.

\vspace{0.2cm}\noindent\textbf{(iii) Fermion-Gauge Boson Interaction:}\vspace{0.1cm}

\noindent The three-loop vertex correction can be written as
\begin{equation}\label{Eq:FFA_3-Loop}
    \begin{aligned}
        i \widetilde{\Gamma}_{\psi\overline{\psi}A}^{ji,\mu} \big|_{\text{div}}^{3} &= 
        - \frac{i}{(16 \pi^2)^3} \, e^7 \, 
        \bigg\{ \mathcal{C}_{\psi\overline{\psi}A}^{3, \, ji} \, \frac{1}{\epsilon^3} + \mathcal{B}_{\psi\overline{\psi}A}^{3, \, ji} \, \frac{1}{\epsilon^2} + \mathcal{A}_{\psi\overline{\psi}A}^{3, \, ji} \, \frac{1}{\epsilon} \bigg\} \, \overline{\gamma}^{\mu} \, \mathbb{P}_{\mathrm{R}}\\
        &= - \frac{i}{(16 \pi^2)^3} \, e^7 \, \big(\mathcal{Y}_R\big)_{jk} \,
        \bigg\{ \mathcal{C}_{\psi\overline{\psi}, \, ki}^{3, \, \text{inv}} \, \frac{1}{\epsilon^3} + \mathcal{B}_{\psi\overline{\psi}, \, ki}^{3, \, \text{inv}} \, \frac{1}{\epsilon^2} + \mathcal{A}_{\psi\overline{\psi}, \, ki}^{3, \, \text{inv}} \, \frac{1}{\epsilon} \bigg\} \, \overline{\gamma}^{\mu} \, \mathbb{P}_{\mathrm{R}},
    \end{aligned}
\end{equation}
with three-loop coefficients (\ref{Eq:CFFA}) to (\ref{Eq:AFFA}).
In the second line the result is by definition completely expressed in terms of the 
invariant coefficients already used for the fermion self energy (\ref{Eq:FermionSelfEnergy_3-Loop}) 
given in Eqs.\ (\ref{Eq:CfermionInv},\ref{Eq:BfermionInv},\ref{Eq:AfermionInv}).
This reflects the convention explained above to attribute the breaking of the relevant 
Ward identity entirely to the fermion self energy.

\vspace{0.2cm}\noindent\textbf{(iv) Triple Gauge Boson Interaction:}\vspace{0.1cm}

\noindent As expected, and as for the one- and two-loop case, the triple gauge boson interaction
does not provide a divergent contribution,
\begin{equation}
    \begin{aligned}
        i \widetilde{\Gamma}_{AAA}^{\rho\nu\mu} \big|_{\text{div}}^{3} = 0.
    \end{aligned}
\end{equation}

\vspace{0.2cm}\noindent\textbf{(v) Quartic Gauge Boson Interaction:}\vspace{0.1cm}

\noindent Unlike in the one- and two-loop case, cf.\ 
Eqs.\ (\ref{Eq:Ssct_1-Loop},\ref{Eq:Ssct_2-Loop}),
in the three-loop case there is a non-evanescent, symmetry-breaking, divergent contribution from
the quartic gauge boson interaction of the form
\begin{equation}
    \begin{aligned}
        i \widetilde{\Gamma}_{AAAA}^{\sigma\rho\nu\mu} \big|_{\text{div}}^{3} &= \frac{i}{(16 \pi^2)^3} \, e^8 \, \overline{\mathcal{A}}_{AAAA}^{3, \, \text{break}} \, \frac{1}{\epsilon} \, \Big( \overline{\eta}^{\mu\nu} \, \overline{\eta}^{\rho\sigma} + \overline{\eta}^{\mu\rho} \, \overline{\eta}^{\nu\sigma} + \overline{\eta}^{\mu\sigma} \, \overline{\eta}^{\nu\rho} \Big),
    \end{aligned}
\end{equation}
with three-loop coefficient (\ref{Eq:AQuarticPhotonBreak}).
This contribution generates a new singular counterterm which first appears at the three-loop level.

\subsection{Three-Loop Breaking of BRST Symmetry}\label{Sec:3-Loop-BRST-Breaking}

We continue with $\Delta$-operator inserted, power-counting UV-divergent, 1PI Green functions, 
from which we obtain the complete BRST breaking at a given loop order, i.e.\ not only the 
divergent symmetry-breaking contributions, but also the finite ones.
The latter is possible due to the usage of the quantum action principle, allowing to
rewrite the symmetry-breaking as an operator insertion, as already explained above.
In particular, we are following the procedure illustrated in Sec.\ \ref{Sec:SymmetryRestoration},
using Eq.\ (\ref{Eq:PerturbativeRequirementAndStartingPoint}) as a starting point.

On the one hand, from the finite contributions, the important finite 
symmetry-restoring counterterms can be derived.
On the other hand, it can be seen that the divergent symmetry-breaking contributions are 
indeed the same as the symmetry-violating contributions already encountered in the last section, 
which serves as a consistency check.

As above, the results are provided in terms of coefficients. 
The coefficients $\mathcal{A}$, $\mathcal{B}$, $\mathcal{C}$ for the divergences 
are the same as the ones used in Sec.\ \ref{Sec:3-Loop-Div-GreenFuncs}, whereas
the coefficients $\mathcal{F}$ are new and correspond to the finite symmetry breakings. 
Their values are given in Eqs.\ 
(\ref{Eq:FPhotonBreak},\ref{Eq:FQuarticPhotonBreak},\ref{Eq:FfermionBreak}).

Note that, although the two 1PI Green functions for the ghost-gauge boson-fermion-fermion
($cA\overline{\psi}\psi$)
and the ghost-quartic gauge boson ($cAAAA$) contributions are both power-counting UV-divergent,
neither of the two Green functions gives rise to a non-vanishing contribution, which is due to
a cancellation of the leading power-counting term in the integrand in the considered abelian theory, 
effectively reducing the power-counting degree by one, 
for all such $\Delta$-operator inserted Green functions.\footnote{
This is true only in the considered abelian theory with the given interaction
structure, cf.\ Sec.\ \ref{Sec:ToDandBeyond}.}
Further, the latter Green function can also not contribute as it could not give rise to 
a renormalisable operator in the
counterterm action, which becomes clear when considering the inverse BRST transformation
of the associated operator, cf.\ Eq.\ (\ref{Eq:BRST-Trafos}). 
In particular, the operator emerges from the BRST transformation 
of $AAAAA$, which is non-renormalisable.\footnote{Note that this changes in non-abelian 
gauge theories with more involved BRST transformations.}

In the following we provide the complete list of results for all relevant 
$\Delta$-operator inserted Green functions.

\vspace{0.2cm}\noindent\textbf{(vi) Ghost-Gauge Boson Contribution:}

\begin{equation}
    \begin{aligned}
        i \bigg( \Big[ \widehat{\Delta} + \Delta_{\text{ct}} \Big] \cdot \widetilde{\Gamma} \bigg)_{A_{\mu} c}^{3} = &- \frac{e^6}{(16 \pi^2)^3} \, \bigg[ \widehat{\mathcal{C}}_{AA}^{3, \text{break}} \, \frac{1}{\epsilon^3} + \widehat{\mathcal{B}}_{AA}^{3, \text{break}} \, \frac{1}{\epsilon^2} + \widehat{\mathcal{A}}_{AA}^{3, \text{break}} \, \frac{1}{\epsilon} \bigg] \, \widehat{p}^{2} \, \overline{p}^{\mu}\\
        &+ \frac{e^6}{(16 \pi^2)^3} \, \bigg[ \overline{\mathcal{A}}_{AA}^{3, \text{break}} \, \frac{1}{\epsilon} + \mathcal{F}_{AA}^{3, \text{break}} \bigg] \, \overline{p}^{2} \, \overline{p}^{\mu},
    \end{aligned}
\end{equation}
with $p$ being the incoming ghost momentum.

\vspace{0.2cm}\noindent\textbf{(vii) Ghost-Fermion-Fermion Contribution:}

\begin{equation}
    \begin{aligned}
        i \bigg( \Big[ \widehat{\Delta} &+ \Delta_{\mathrm{ct}} \Big] \cdot \widetilde{\Gamma} \bigg)_{\psi_j \overline{\psi}_i c}^{3}\\
        &= - \frac{e^7}{(16 \pi^2)^3} \, \big(\mathcal{Y}_R\big)_{jk} \, 
        \bigg\{  \mathcal{B}_{\psi\overline{\psi}, \, ki}^{3, \, \text{break}} \, \frac{1}{\epsilon^2} + \mathcal{A}_{\psi\overline{\psi}, \, ki}^{3, \, \text{break}} \, \frac{1}{\epsilon} + \mathcal{F}_{\psi\overline{\psi}, \, ki}^{3, \text{break}} \bigg\} \, \Big( \overline{\slashed{p}}_1 + \overline{\slashed{p}}_2 \Big) \, \mathbb{P}_{\mathrm{R}},
    \end{aligned}
\end{equation}
with $p_1$ and $p_2$ being the incoming fermion momenta.

\vspace{0.2cm}\noindent\textbf{(viii) Ghost-double Gauge Boson Contribution:}

\begin{equation}
    \begin{aligned}
        &i \bigg( \Big[ \widehat{\Delta} + \Delta_{\text{ct}} \Big] \cdot \widetilde{\Gamma} \bigg)_{A_{\nu}A_{\mu}c}^{3} \propto \, \text{Tr}\big(\mathcal{Y}_R^3\big) \, \varepsilon^{\mu\nu\rho\sigma} = 0,
    \end{aligned}
\end{equation}
which vanishes identically due to the anomaly cancellation condition (\ref{Eq:AnomalyCancellationCondition}),
used here and in all other Green functions.

\vspace{0.2cm}\noindent\textbf{(ix) Ghost-triple Gauge Boson Contribution:}

\begin{equation}
    \begin{aligned}
        i \bigg( \Big[ \widehat{\Delta} + \Delta_{\text{ct}} \Big] \cdot \widetilde{\Gamma} \bigg)_{A_{\rho}A_{\nu}A_{\mu}c}^{3} = &- \frac{e^8}{(16 \pi^2)^3} \, \bigg[ \overline{\mathcal{A}}_{AAAA}^{3, \, \text{break}} \, \frac{1}{\epsilon} + \mathcal{F}_{AAAA}^{3, \, \text{break}} \bigg]\\
        &\times \, \big(\overline{p}_1 + \overline{p}_2 + \overline{p}_3\big)_{\sigma} \, \Big( \overline{\eta}^{\mu\nu} \, \overline{\eta}^{\rho\sigma} + \overline{\eta}^{\mu\rho} \, \overline{\eta}^{\nu\sigma} + \overline{\eta}^{\mu\sigma} \, \overline{\eta}^{\nu\rho} \Big),
    \end{aligned}
\end{equation}
with $p_1$, $p_2$ and $p_3$ being the incoming gauge boson momenta.

Ultimately, the full breaking of the Slavnov-Taylor identity at the three-loop level
reads
\begin{equation}
    \begin{aligned}
        \Big( \big[ \widehat{\Delta} + \Delta_{\mathrm{ct}} \big] &\cdot \widetilde{\Gamma} \Big)^{3}\\
        = &- \frac{e^6}{(16 \pi^2)^3} \,
        \bigg[ \widehat{\mathcal{C}}_{AA}^{3, \text{break}} \, \frac{1}{\epsilon^3} + \widehat{\mathcal{B}}_{AA}^{3, \text{break}} \, \frac{1}{\epsilon^2} + \widehat{\mathcal{A}}_{AA}^{3, \text{break}} \, \frac{1}{\epsilon} \bigg]
        \Dintx \, c \, \overline{\partial}_{\mu} \, \widehat{\partial}^2 \, \overline{A}^{\mu}\\
        &+ \frac{e^6}{(16 \pi^2)^3} \,
        \bigg[ \overline{\mathcal{A}}_{AA}^{3, \text{break}} \, \frac{1}{\epsilon} + \mathcal{F}_{AA}^{3, \text{break}} \bigg]
        \Dintx \, c \, \overline{\partial}_{\mu} \, \overline{\partial}^2 \, \overline{A}^{\mu}\\
        &- \frac{e^7}{(16 \pi^2)^3} \, \big(\mathcal{Y}_R\big)_{jk}
        \bigg[  \mathcal{B}_{\psi\overline{\psi}, \, ki}^{3, \, \text{break}} \, \frac{1}{\epsilon^2} + \mathcal{A}_{\psi\overline{\psi}, \, ki}^{3, \, \text{break}} \, \frac{1}{\epsilon} + \mathcal{F}_{\psi\overline{\psi}, \, ki}^{3, \, \text{break}} \bigg]\\
        &\hspace{3cm} \times 
        \Dintx \, c \, \overline{\partial}_{\mu} \Big( \overline{\psi}_j \, \overline{\gamma}^{\mu} \,  \mathbb{P}_{\mathrm{R}} \, \psi_i \Big)\\
        &- \frac{e^8}{(16 \pi^2)^3} \,
        \bigg[ \overline{\mathcal{A}}_{AAAA}^{3, \, \text{break}} \, \frac{1}{\epsilon} + \mathcal{F}_{AAAA}^{3, \, \text{break}} \bigg]
        \Dintx \, \frac{1}{2} \, c \, \overline{\partial}_{\mu} \Big( \overline{A}^{\mu} \overline{A}_{\nu} \overline{A}^{\nu} \Big) + \mathcal{O}(\hat{.})
    \end{aligned}
\end{equation}

\subsection{Three-Loop Singular Counterterm Action}

Combining all results, the complete singular counterterm action at the three-loop level is given by
\begin{equation}\label{Eq:Ssct_3-Loop}
    \begin{aligned}
        S^{3}_{\mathrm{sct}} = 
        &\phantom{-\,\,} \frac{e^6}{(16 \pi^2)^3} \,
        \bigg[ \mathcal{B}_{AA}^{3, \text{inv}} \, \frac{1}{\epsilon^2} + \mathcal{A}_{AA}^{3, \text{inv}} \, \frac{1}{\epsilon} \bigg] 
         \, \Dintx \, \Big(-\frac{1}{4} \, \overline{F}^{\mu\nu} \, \overline{F}_{\mu\nu}\Big)\\
        &- \frac{e^6}{(16 \pi^2)^3} \, 
        \bigg[ \mathcal{C}_{\psi\overline{\psi}, \, ji}^{3, \, \text{inv}} \, \frac{1}{\epsilon^3} + \mathcal{B}_{\psi\overline{\psi}, \, ji}^{3, \, \text{inv}} \, \frac{1}{\epsilon^2} + \mathcal{A}_{\psi\overline{\psi}, \, ji}^{3, \, \text{inv}} \, \frac{1}{\epsilon} \bigg] \\
        &\hspace{1.75cm}\times \Dintx \, \Big( \overline{\psi}_j \, i \, \overline{\slashed{\partial}} \,  \mathbb{P}_{\mathrm{R}} \, \psi_i - e \,  \big(\mathcal{Y}_R\big)_{kj} \, \overline{\psi}_k \, \overline{\slashed{A}} \, \mathbb{P}_{\mathrm{R}} \, \psi_i \Big)\\
        &- \frac{e^6}{(16 \pi^2)^3} \,
        \bigg[ \widehat{\mathcal{C}}_{AA}^{3, \text{break}} \, \frac{1}{\epsilon^3} + \widehat{\mathcal{B}}_{AA}^{3, \text{break}} \, \frac{1}{\epsilon^2} + \widehat{\mathcal{A}}_{AA}^{3, \text{break}} \, \frac{1}{\epsilon} \bigg] \,
        \Dintx \, \frac{1}{2} \, \overline{A}_{\mu} \, \widehat{\partial}^2 \, \overline{A}^{\mu}\\
        &+ \frac{e^6}{(16 \pi^2)^3} \,
        \overline{\mathcal{A}}_{AA}^{3, \text{break}} \, \frac{1}{\epsilon} \,
        \Dintx \, \frac{1}{2} \, \overline{A}_{\mu} \, \overline{\partial}^2 \, \overline{A}^{\mu}\\
        &- \frac{e^6}{(16 \pi^2)^3} \, 
        \bigg[ \mathcal{B}_{\psi\overline{\psi}, \, ji}^{3, \, \text{break}} \, \frac{1}{\epsilon^2} + \mathcal{A}_{\psi\overline{\psi}, \, ji}^{3, \, \text{break}} \, \frac{1}{\epsilon} \bigg] \,
        \Dintx \, \Big( \overline{\psi}_j \, i \, \overline{\slashed{\partial}} \,  \mathbb{P}_{\mathrm{R}} \, \psi_i \Big)\\
        &- \frac{e^8}{(16 \pi^2)^3} \,
        \overline{\mathcal{A}}_{AAAA}^{3, \, \text{break}} \, \frac{1}{\epsilon} \,
        \Dintx \, \frac{1}{8} \, \overline{A}_{\mu} \, \overline{A}^{\mu} \, \overline{A}_{\nu} \, \overline{A}^{\nu}.
    \end{aligned}
\end{equation}
Including these counterterms removes all three-loop UV-divergences from the theory, and together with
the one- and two-loop counterterms in Eq.\ (\ref{Eq:Ssct_1-Loop}) and Eq.\ (\ref{Eq:Ssct_2-Loop}),
respectively, they guarantee a finite theory up to the three-loop level.
The first two lines of Eq.\ (\ref{Eq:Ssct_3-Loop}) represent the BRST invariant piece of the
counterterm action, which could be written as parameter and field renormalisation; the rest are singular BRST breaking contributions.
We highlight that there are two kinds of changes compared  to the one- and two-loop case given in Eqs.\ 
(\ref{Eq:Ssct_1-Loop},\ref{Eq:Ssct_2-Loop}). First, there are  higher order $\epsilon$-poles for
already earlier appearing counterterm structures. Second and more interestingly, there are two
completely new counterterms generated at the three-loop level:
the non-evanescent bilinear gauge boson counterterm in the fourth line and
the non-evanescent quartic gauge boson counterterm in the last line of
Eq.\ (\ref{Eq:Ssct_3-Loop}).
Beyond that and following on from the previous discussion below Eqs.\ 
(\ref{Eq:FermionSelfEnergy_3-Loop}) and (\ref{Eq:FFA_3-Loop});
here, it becomes very clear that the choice that only bilinear fermion
terms contribute to the BRST breaking part, see the penultimate line of Eq.\ (\ref{Eq:Ssct_3-Loop}),
and fermion-gauge boson interaction terms do not is not unique.
We could have also chosen it vice versa.

\subsection{Three-Loop Finite Symmetry-Restoring Counterterm Action}

Finally, the full three-loop finite symmetry-restoring counterterm action takes the form
\begin{equation}
    \begin{aligned}
        S^{3}_{\mathrm{fct}} = &\phantom{+\,\,} \frac{e^6}{(16 \pi^2)^3} \,
        \mathcal{F}_{AA}^{3, \text{break}}
        \intx \, \frac{1}{2} \, \overline{A}_{\mu} \, \overline{\partial}^2 \, \overline{A}^{\mu}
        - \frac{e^6}{(16 \pi^2)^3} \, 
        \mathcal{F}_{\psi\overline{\psi}, \, ji}^{3, \, \text{break}}
        \intx \, \overline{\psi}_j \, i \, \overline{\slashed{\partial}} \, \mathbb{P}_{\mathrm{R}} \, \psi_i\\
        &- \frac{e^8}{(16 \pi^2)^3} \,
        \mathcal{F}_{AAAA}^{3, \, \text{break}}
        \intx \, \frac{1}{8} \, \overline{A}_{\mu} \overline{A}^{\mu} \overline{A}_{\nu} \overline{A}^{\nu}.
    \end{aligned}
\end{equation}
Together with the one- and two-loop counterterms in Eq.\ (\ref{Eq:Sfct_1-Loop}) 
and Eq.\ (\ref{Eq:Sfct_2-Loop}), these counterterms guarantee
that the theory satisfies the 
Slavnov-Taylor identity after renormalisation up to the three-loop level.
In contrast to the singular counterterm action there are no new counterterm structures
emerging at the three-loop level.
There are still the same three counterterms as in the one- and two-loop case,
cf.\ Eqs.\ (\ref{Eq:Sfct_1-Loop},\ref{Eq:Sfct_2-Loop}),
just with different coefficients. They correspond to the restoration of the transversality of the 
gauge boson self energy, the Ward identity between the fermion self energy and the fermion-gauge boson
three-point function, and the Ward identity for the quartic gauge boson self interaction.
The reason for the simplicity of these counterterms is that the symmetry-restoring counterterms 
may be defined purely in 4 dimensions and
are restricted by power-counting. Hence, we also 
expect the same counterterm structure to continue to higher loop levels.


\section{Conclusion}\label{Sec:Conclusion}

In this work, we successfully performed the complete three-loop renormalisation of an abelian 
chiral gauge theory within the framework of DReg, treating $\gamma_5$ rigorously 
as a non-anticommuting object in the BMHV scheme.
In particular, we computed not only the singular, but also the complete finite 
symmetry-restoring counterterm action up to the three-loop level.
While the first is necessary to render the theory finite, the latter is needed to cancel the spurious
symmetry-breaking induced by the BMHV scheme, such that the renormalised theory is both finite 
and gauge invariant.

Technically we employed an efficient procedure for the symmetry restoration developed and applied 
before at the one-loop and two-loop order to chiral gauge theories. Using the quantum action principle 
the symmetry-breaking can be obtained from Green functions with evanescent operator insertions, 
and the Slavnov-Taylor identity acts as a symmetry requirement for the determination of 
symmetry-restoring counterterms
at any given loop order.
The efficiency of this method stems from the fact that only the UV-divergent part of power-counting
divergent Green functions needs to be calculated to obtain all necessary counterterms, including
the finite symmetry-restoring ones.
In this work, we have now successfully applied this procedure at the three-loop level.
To this end, we have upgraded our computational setup and implemented the so-called
all massive tadpoles method. This method represents an infrared rearrangement and maps
all Feynman diagrams to fully massive single-scale vacuum bubbles, such that all UV-divergences
can ultimately be extracted from solving tadpole master integrals.

In the singular counterterm action we encountered not only higher order $\epsilon$-poles 
for already earlier appearing BRST breaking counterterm structures emerging at the
three-loop level, but also new counterterm structures were generated at 
the three-loop level for the first time.
In particular, a new non-evanescent bilinear gauge boson counterterm and a new non-evanescent
quartic gauge boson counterterm, both UV-divergent and BRST-breaking, emerge for the first time
at the three-loop level.
In contrast to this, in the finite symmetry-restoring counterterm action, there are no new 
counterterm structures emerging at the three-loop level.
These admit still the same counterterm structures as in the one- and two-loop case, just with 
different coefficients.
As a matter of principle, both the singular and the finite counterterms are restricted by power counting. 
But the singular counterterms can involve 
$D$-, $4$- or $(-2\epsilon)$-dimensional Lorentz covariants and thus a larger number of different structures,
whereas the finite symmetry-restoring counterterms may be defined in purely 4 dimensions and thus involve 
only a small number of different structures. Our three-loop findings are in line with these general statements.

We have shown that $\gamma_5$  can be treated rigorously and systematically at high loop orders in the 
context of chiral gauge theories, 
without any ambiguities or the need for external arguments, by using the BMHV scheme.
Although the singular counterterm action obtains new contributions, the counterterm action can still
be written in a rather compact form, suitable for computer implementations.
This is crucial and becomes necessary in future calculations of electroweak precision observables 
in order to achieve the required higher precision to align with the increasing experimental precision.

We were able to further automate our methodology, such that calculations in other theories, 
and at even higher loop orders come in reach.
Most importantly, a consistent renormalisation of the Standard Model
at the multi-loop level employing the BMHV scheme will become possible using the methods of the present paper.


\section*{Acknowledgments} 

D.S.\ and M.W.\ acknowledge financial support by the German Science 
Foundation DFG, grant STO 876/8-1. 
In particular, we would like to thank Vladyslav Shtabovenko not 
only for maintaining and developing the software
\texttt{FeynCalc} and \texttt{FeynHelpers}, but also for many useful discussions that 
helped in developing the computational setup for evaluating the Feynman diagrams in this 
research work.
We also want to thank our collaborators Herm\`es B\'elusca-Ma\"\i{}to, Amon Ilakovac, 
Paul Kühler and Marija Ma\dj{}or-Bo\v{z}inovi\'c for insightful ideas 
and valuable discussions. 
Further, we would like to thank Konstantin Chetyrkin, Michal Czakon, 
Andreas Maier, Andreas von Manteuffel, Peter Marquard, York Schröder and Max Zoller for 
useful discussions and intellectual exchange. We are grateful to the Centre for 
Information Services and High Performance Computing $[$Zentrum für Informationsdienste und 
Hochleistungsrechnen (ZIH)$]$ TU Dresden for providing its facilities for high-performance calculations.

\begin{appendices}
\addappheadtotoc

\section{Explicit Results for the Three-Loop Coefficients} \label{App:3-Loop-Coeffs}

In this section of the appendix we provide explicit results for the three-loop coefficients
used in Sec.\ \ref{Sec:3-Loop-Ren-Results}.
We begin with the coefficients for the purely gauge bosonic terms:

\vspace{0.3cm}\noindent\textbf{Gauge Boson Three-Loop Coefficients:}

\begin{align}
    \mathcal{B}_{AA}^{3, \text{inv}} &= \frac{4}{162} \Big( 3 \, \text{Tr}\big(\mathcal{Y}_R^6\big) - 5 \, \text{Tr}\big(\mathcal{Y}_R^4\big) \text{Tr}\big(\mathcal{Y}_R^2\big) \Big)\label{Eq:BphotonInv}\\
    \mathcal{A}_{AA}^{3, \text{inv}} &= - \frac{1}{1620} \Big( 2552 \, \text{Tr}\big(\mathcal{Y}_R^6\big) + 61 \, \text{Tr}\big(\mathcal{Y}_R^4\big) \text{Tr}\big(\mathcal{Y}_R^2\big) \Big)\label{Eq:AphotonInv}
\end{align}

\begin{align}
    \widehat{\mathcal{C}}_{AA}^{3, \text{break}} &= \frac{1}{18} \, \text{Tr}\big(\mathcal{Y}_R^6\big)\label{Eq:CphotonBreakEvan}\\
    \widehat{\mathcal{B}}_{AA}^{3, \text{break}} &= - \frac{1}{1080} \Big( 529 \, \text{Tr}\big(\mathcal{Y}_R^6\big) + 122 \, \text{Tr}\big(\mathcal{Y}_R^4\big) \text{Tr}\big(\mathcal{Y}_R^2 \Big)\label{Eq:BphotonBreakEvan}\\
    \widehat{\mathcal{A}}_{AA}^{3, \text{break}} &= \frac{1}{64800} \Big( \big( 156672 \, \zeta_{3} - 49427 \big) \text{Tr}\big(\mathcal{Y}_R^6\big) - 8374 \, \text{Tr}\big(\mathcal{Y}_R^4\big) \text{Tr}\big(\mathcal{Y}_R^2\big) \Big)\label{Eq:AphotonBreakEvan}
\end{align}

\begin{align}
    \overline{\mathcal{A}}_{AA}^{3, \text{break}} &= \frac{1}{1080} \Big( 18 \, \text{Tr}\big(\mathcal{Y}_R^6\big) + 79 \, \text{Tr}\big(\mathcal{Y}_R^4\big) \text{Tr}\big(\mathcal{Y}_R^2\big) \Big)\label{Eq:AphotonBreakNonEvan}
\end{align}

\begin{align}\label{Eq:FPhotonBreak}
    \mathcal{F}_{AA}^{3, \text{break}} &= - \frac{1}{21600} \Big( \big( 35242 + 8448 \, \zeta_{3} \big) \text{Tr}\big(\mathcal{Y}_R^6\big) + 1639 \, \text{Tr}\big(\mathcal{Y}_R^4\big) \text{Tr}\big(\mathcal{Y}_R^2\big) \Big)
\end{align}

\begin{align}\label{Eq:AQuarticPhotonBreak}
    \overline{\mathcal{A}}_{AAAA}^{3, \text{break}} &= \frac{1}{54} \Big( 6 \, \text{Tr}\big(\mathcal{Y}_R^8\big) + 13 \, \text{Tr}\big(\mathcal{Y}_R^6\big) \text{Tr}\big(\mathcal{Y}_R^2\big) + 48 \, \big(\text{Tr}\big(\mathcal{Y}_R^4\big)\big)^2 \Big)
\end{align}

\begin{align}\label{Eq:FQuarticPhotonBreak}
    \mathcal{F}_{AAAA}^{3, \text{break}} &= -\frac{1}{54} \bigg( \frac{1387+2592\,\zeta_{3}}{10} \, \text{Tr}\big(\mathcal{Y}_R^8\big) +  \frac{101}{20} \, \text{Tr}\big(\mathcal{Y}_R^6\big) \text{Tr}\big(\mathcal{Y}_R^2\big) + 51 \, \big(\text{Tr}\big(\mathcal{Y}_R^4\big)\big)^2 \bigg)
\end{align}
Continuing with the coefficients for terms that contain fermions:

\vspace{0.3cm}\noindent\textbf{Fermion Three-Loop Coefficients:}

\begin{align}
    \mathcal{C}_{\overline{\psi}\psi, \, ij}^{3, \, \text{inv}} &= \frac{1}{6} \big(\mathcal{Y}_R^6\big)_{ij}\label{Eq:CfermionInv}\\
    \begin{split}\label{Eq:BfermionInv}
    \mathcal{B}_{\overline{\psi}\psi, \, ij}^{3, \, \text{inv}} &= \frac{1}{324} \Big( 432 \big(\mathcal{Y}_R^6\big)_{ij} - 186 \big(\mathcal{Y}_R^4\big)_{ij} \text{Tr}\big(\mathcal{Y}_R^2\big)\\
    &\hspace{1.3cm} - 6 \big(\mathcal{Y}_R^2\big)_{ij} \text{Tr}\big(\mathcal{Y}_R^4\big) - \big(\mathcal{Y}_R^2\big)_{ij} \big(\text{Tr}\big(\mathcal{Y}_R^2\big)\big)^2 \Big)
    \end{split}\\
    \begin{split}\label{Eq:AfermionInv}
    \mathcal{A}_{\overline{\psi}\psi, \, ij}^{3, \, \text{inv}} &= \frac{1}{3888} \bigg[ 21843 \big(\mathcal{Y}_R^6\big)_{ij} - 4338 \big(\mathcal{Y}_R^4\big)_{ij} \text{Tr}\big(\mathcal{Y}_R^2\big)\\
    &\hspace{1.3cm} - \Big( 2166 \text{Tr}\big(\mathcal{Y}_R^4\big) - 91 \big(\text{Tr}\big(\mathcal{Y}_R^2\big)\big)^2 \Big) \big(\mathcal{Y}_R^2\big)_{ij} + 2430 \text{Tr}\big(\mathcal{Y}_R^5\big) \big(\mathcal{Y}_R\big)_{ij} \bigg]
    \end{split}
\end{align}

\begin{align}
    \mathcal{B}_{\overline{\psi}\psi, \, ij}^{3, \, \text{break}} &= - \frac{1}{3} \bigg[ \big(\mathcal{Y}_R^6\big)_{ij} - \frac{1}{2} \big(\mathcal{Y}_R^4\big)_{ij} \text{Tr}\big(\mathcal{Y}_R^2\big) + \frac{\big(\mathcal{Y}_R^2\big)_{ij}}{54} \Big( 3 \text{Tr}\big(\mathcal{Y}_R^4\big) + 13 \big(\text{Tr}\big(\mathcal{Y}_R^2\big)\big)^2 \Big) \bigg]\label{Eq:BfermionBreak}\\
    \begin{split}\label{Eq:AfermionBreak}
    \mathcal{A}_{\overline{\psi}\psi, \, ij}^{3, \, \text{break}} &= - \frac{1}{18} \bigg[ 79 \big(\mathcal{Y}_R^6\big)_{ij} - \frac{169}{6} \, \big(\mathcal{Y}_R^4\big)_{ij} \text{Tr}\big(\mathcal{Y}_R^2\big)\\
    &\hspace{1.3cm} - \frac{\big(\mathcal{Y}_R^2\big)_{ij}}{108} \Big( 159 \text{Tr}\big(\mathcal{Y}_R^4\big) - 113 \big(\text{Tr}\big(\mathcal{Y}_R^2\big)\big)^2 \Big) + \frac{45}{4} \, \big(\mathcal{Y}_R\big)_{ij} \text{Tr}\big(\mathcal{Y}_R^5\big) \bigg]
    \end{split}
\end{align}

\begin{align}\label{Eq:FfermionBreak}
    \begin{split}
    \mathcal{F}_{\overline{\psi}\psi, \, ij}^{3, \, \text{break}} = &- \bigg( \frac{775}{54} + \frac{58}{9} \, \zeta_{3} \bigg) \big(\mathcal{Y}_R^6\big)_{ij} + \frac{10}{9} \, \big(\mathcal{Y}_R^4\big)_{ij} \text{Tr}\big(\mathcal{Y}_R^2\big)\\
    &- \big(\mathcal{Y}_R^2\big)_{ij} \Bigg[ \bigg( \frac{9725}{3888} + \frac{14}{3} \, \zeta_{3} \bigg) \text{Tr}\big(\mathcal{Y}_R^4\big) - \frac{1993}{23328} \, \big(\text{Tr}\big(\mathcal{Y}_R^2\big)\big)^2 \Bigg]\\
    &+ \big(\mathcal{Y}_R\big)_{ij} \bigg( \frac{215}{96} - 7 \, \zeta_3 \bigg) \text{Tr}\big(\mathcal{Y}_R^5\big)
    \end{split}
\end{align}

\noindent Finally, some relations among certain coefficients are in place:

\begin{align}
    \mathcal{C}_{\overline{\psi}\psi}^{3, \, ij} &= \mathcal{C}_{\overline{\psi}\psi, \, ij}^{3, \, \text{inv}}\\
    \mathcal{B}_{\overline{\psi}\psi}^{3, \, ij} &= \mathcal{B}_{\overline{\psi}\psi, \, ij}^{3, \, \text{inv}} + \mathcal{B}_{\overline{\psi}\psi, \, ij}^{3, \, \text{break}}\\
    \mathcal{A}_{\overline{\psi}\psi}^{3, \, ij} &= \mathcal{A}_{\overline{\psi}\psi, \, ij}^{3, \, \text{inv}} + \mathcal{A}_{\overline{\psi}\psi, \, ij}^{3, \, \text{break}}
\end{align}

\begin{align}
    \mathcal{C}_{\overline{\psi}A\psi}^{3, \, ij} &= 
    \big(\mathcal{Y}_R\big)_{ik}\,\,\mathcal{C}_{\overline{\psi}\psi, \, kj}^{3, \, \text{inv}}
    \label{Eq:CFFA}\\
    \mathcal{B}_{\overline{\psi}A\psi}^{3, \, ij} &= 
    \big(\mathcal{Y}_R\big)_{ik}\,\,\mathcal{B}_{\overline{\psi}\psi, \, kj}^{3, \, \text{inv}}
    \label{Eq:BFFA}\\
    \mathcal{A}_{\overline{\psi}A\psi}^{3, \, ij} &= 
    \big(\mathcal{Y}_R\big)_{ik}\,\,\mathcal{A}_{\overline{\psi}\psi, \, kj}^{3, \, \text{inv}}
    \label{Eq:AFFA}
\end{align}


\section{One-Loop Results} \label{App:1-Loop-Results}

Here, we provide the complete results for a full one-loop renormalisation of the
considered abelian chiral gauge theory in $R_{\xi}$-gauge. We find perfect agreement with
Ref.\ \cite{Belusca-Maito:2021lnk}, up to a different sign convention in the covariant derivative as
already stated before, cf.\ Sec.\ \ref{Sec:ToDandBeyond}.

First, the full one-loop breaking of the Slavnov-Taylor identity is given by
\begin{equation}
    \begin{aligned}
        \big( \widehat{\Delta} \cdot \widetilde{\Gamma} \big)^{1}
        = &- \frac{1}{16 \pi^2} \, \Dintx \, \bigg\{ \frac{e^2 \, \text{Tr}\big(\mathcal{Y}_R^2\big)}{3} \, \bigg[ \frac{1}{\epsilon} \, c \, \overline{\partial}_{\mu} \, \widehat{\partial}^2 \, \overline{A}^{\mu} + c \, \overline{\partial}_{\mu} \, \overline{\partial}^2 \, \overline{A}^{\mu} \bigg]\\
        &- \frac{5 + \xi}{6} \, e^3 \, \big(\mathcal{Y}_R^3\big)_{ji} \, c \, \overline{\partial}_{\mu} \Big( \overline{\psi}_j \, \overline{\gamma}^{\mu} \,  \mathbb{P}_{\mathrm{R}} \, \psi_i \Big) - \frac{e^4 \, \text{Tr}\big(\mathcal{Y}_R^4\big)}{3} \, c \, \overline{\partial}_{\mu} \Big( \overline{A}^{\mu} \overline{A}_{\nu} \overline{A}^{\nu} \Big)\\
        &+ \mathcal{O}(\hat{.}) \bigg\}.
    \end{aligned}
\end{equation}
Eventually, the one-loop singular counterterm action takes the form
\begin{equation}\label{Eq:Ssct_1-Loop}
    \begin{aligned}
        S^{1}_{\mathrm{sct}} = &- \frac{e^2}{16 \pi^2} \, \frac{1}{\epsilon} \, \bigg[ \frac{2}{3} \,  \text{Tr}\big(\mathcal{Y}_R^2\big) \Dintx \, \Big(-\frac{1}{4} \, \overline{F}^{\mu\nu} \, \overline{F}_{\mu\nu}\Big)\\
        &\hspace{1.7cm} + \xi \, \big(\mathcal{Y}_R^2\big)_{ji} 
        \Dintx \, \Big( \overline{\psi}_j \, i \, \overline{\slashed{\partial}} \,  \mathbb{P}_{\mathrm{R}} \, \psi_i - e \,  \big(\mathcal{Y}_R\big)_{kj} \, \overline{\psi}_k \, \overline{\slashed{A}} \, \mathbb{P}_{\mathrm{R}} \, \psi_i \Big)\\
        &\hspace{1.7cm} + \frac{\text{Tr}\big(\mathcal{Y}_R^2\big)}{3} \Dintx \, \frac{1}{2} \, \overline{A}_{\mu} \, \widehat{\partial}^2 \, \overline{A}^{\mu} \bigg],
    \end{aligned}
\end{equation}
whereas the one-loop finite symmetry-restoring counterterm action can be written as
\begin{equation}\label{Eq:Sfct_1-Loop}
    \begin{aligned}
        S^{1}_{\mathrm{fct}} = &- \frac{1}{16 \pi^2} \intx \bigg[ \frac{e^2}{3} \, \text{Tr}\big(\mathcal{Y}_R^2\big) \, \frac{1}{2} \, \overline{A}_{\mu} \, \overline{\partial}^2 \, \overline{A}^{\mu} 
        - \frac{2 \, e^4}{3} \, \text{Tr}\big(\mathcal{Y}_R^4\big) \, \frac{1}{8} \, \overline{A}_{\mu} \overline{A}^{\mu} \overline{A}_{\nu} \overline{A}^{\nu}\\
        &\hspace{2.4cm} - \frac{5+\xi}{6} \, e^2 \, \big(\mathcal{Y}_R^2\big)_{ji} \, \overline{\psi}_j \, i \, \overline{\slashed{\partial}} \, \mathbb{P}_{\mathrm{R}} \, \psi_i \bigg].
    \end{aligned}
\end{equation}


\section{Two-Loop Results} \label{App:2-Loop-Results}

Finally, we provide the complete results for a full two-loop renormalisation of the
considered abelian chiral gauge theory in $R_{\xi}$-gauge.
In contrast to the one-loop results they have been published only in Feynman gauge,
i.e.\ $\xi=1$, in Ref.\ \cite{Belusca-Maito:2021lnk} so far.

In the limit $\xi=1$ we again find perfect agreement with
Ref.\ \cite{Belusca-Maito:2021lnk}, up to the different sign convention 
in the covariant derivative and a typo in Ref.\ \cite{Belusca-Maito:2021lnk}
in the ghost-triple gauge boson term in the breaking of the Slavnov-Taylor identity
and equivalently in the quartic gauge boson term of the finite symmetry-restoring counterterm action,
cf.\ the last term in Eq.\ (\ref{Eq:2-Loop-STIBreaking}) and the second term in 
Eq.\ (\ref{Eq:Sfct_2-Loop}), respectively.
In Ref.\ \cite{Belusca-Maito:2021lnk}, there is a factor of $-1/2$ missing, which we have corrected
here.

With this being said, the full two-loop breaking of the Slavnov-Taylor identity reads
    \begin{align}\label{Eq:2-Loop-STIBreaking}
        \Big( \big[ \widehat{\Delta} &+ \Delta_{\mathrm{ct}} \big] \cdot \widetilde{\Gamma} \Big)^{2}\nonumber\\
        = &- \frac{1}{(16 \pi^2)^2} \, \Dintx \, \bigg\{ \frac{e^4 \, \text{Tr}\big(\mathcal{Y}_R^4\big)}{6} \, \bigg[ \bigg(\frac{\xi}{\epsilon^2} - \frac{43-26\,\xi}{12}\frac{1}{\epsilon}\bigg) c \, \overline{\partial}_{\mu} \, \widehat{\partial}^2 \, \overline{A}^{\mu} - \frac{5\,\xi+17}{8} c \, \overline{\partial}_{\mu} \, \overline{\partial}^2 \, \overline{A}^{\mu} \bigg]\nonumber\\
        &+ \frac{e^5}{3} \, \bigg[ \bigg( \frac{3\,\xi+17}{8}\,\xi\, \big(\mathcal{Y}_R^5\big)_{ji} - \frac{3\,\xi^2+4\,\xi+153}{240} \, \text{Tr}\big(\mathcal{Y}_R^2\big) \big(\mathcal{Y}_R^3\big)_{ji} \bigg) \frac{1}{\epsilon}\nonumber\\
        &\hspace{0.9cm} + \frac{3\,\xi^2+519\,\xi+4558}{480}\,\big(\mathcal{Y}_R^5\big)_{ji} - \frac{471\,\xi^2-92\,\xi+1221}{14400} \, \text{Tr}\big(\mathcal{Y}_R^2\big) \big(\mathcal{Y}_R^3\big)_{ji} \bigg]\\
        &\hspace{0.9cm} \times \, c \, \overline{\partial}_{\mu} \Big( \overline{\psi}_j \, \overline{\gamma}^{\mu} \,  \mathbb{P}_{\mathrm{R}} \, \psi_i \Big)\nonumber\\
        &+ \frac{3\,e^6\,\text{Tr}\big(\mathcal{Y}_R^6\big)}{4} \, \frac{\xi+5}{6} \, c \, \overline{\partial}_{\mu} \Big( \overline{A}^{\mu} \overline{A}_{\nu} \overline{A}^{\nu} \Big) + \mathcal{O}(\hat{.}) \bigg\}.\nonumber
    \end{align}
The two-loop singular counterterm action can be written as
\begin{equation}\label{Eq:Ssct_2-Loop}
    \begin{aligned}
        S^{2}_{\mathrm{sct}} = &- \frac{e^4}{(16 \pi^2)^2} \, \frac{2\,\text{Tr}\big(\mathcal{Y}_R^4\big)}{3} \, \frac{2+\xi}{3} \, \frac{1}{\epsilon} \, \Dintx \, \Big(-\frac{1}{4} \, \overline{F}^{\mu\nu} \, \overline{F}_{\mu\nu}\Big)\\
        &+ \frac{e^4}{(16 \pi^2)^2} \bigg[ \frac{\xi^2}{2} \, \big(\mathcal{Y}_R^4\big)_{ji} \, \frac{1}{\epsilon^2} + \bigg( \frac{9(1+\xi)-\xi^2}{12} \big(\mathcal{Y}_R^4\big)_{ji}\\
        &\hspace{4.1cm} - \frac{24\xi^2-3\xi-1}{20} \, \frac{\text{Tr}\big(\mathcal{Y}_R^2\big)}{9} \big(\mathcal{Y}_R^2\big)_{ji}\bigg) \, \frac{1}{\epsilon} \bigg]\\
        &\hspace{1.75cm} \times \Dintx \, \Big( \overline{\psi}_j \, i \, \overline{\slashed{\partial}} \,  \mathbb{P}_{\mathrm{R}} \, \psi_i - e \,  \big(\mathcal{Y}_R\big)_{kj} \, \overline{\psi}_k \, \overline{\slashed{A}} \, \mathbb{P}_{\mathrm{R}} \, \psi_i \Big)\\
        &- \frac{e^4}{(16 \pi^2)^2} \, \frac{\text{Tr}\big(\mathcal{Y}_R^4\big)}{3} \bigg[\frac{\xi}{4}\,\frac{1}{\epsilon^2} - \frac{43-26\xi}{48}\frac{1}{\epsilon}\bigg] \Dintx \, \frac{1}{2} \, \overline{A}_{\mu} \, \widehat{\partial}^2 \, \overline{A}^{\mu}\\
        &- \frac{e^4}{(16 \pi^2)^2} \bigg[ \frac{\xi(17+3\xi)}{24} \big(\mathcal{Y}_R^4\big)_{ji} - \frac{153+4\xi+3\xi^2}{12} \, \frac{\text{Tr}\big(\mathcal{Y}_R^2\big)}{60} \big(\mathcal{Y}_R^2\big)_{ji}\bigg] \, \frac{1}{\epsilon}\\
        &\hspace{1.75cm} \times
        \Dintx \, \Big( \overline{\psi}_j \, i \, \overline{\slashed{\partial}} \,  \mathbb{P}_{\mathrm{R}} \, \psi_i \Big),
    \end{aligned}
\end{equation}
whereas the two-loop finite symmetry-restoring counterterm action again admits the following structure
\begin{equation}\label{Eq:Sfct_2-Loop}
    \begin{aligned}
        S^{2}_{\mathrm{fct}} &= \frac{1}{(16 \pi^2)^2} \intx \bigg[ \frac{5\xi+17}{48} \, e^4 \, \text{Tr}\big(\mathcal{Y}_R^4\big) \, \frac{1}{2} \, \overline{A}_{\mu} \, \overline{\partial}^2 \, \overline{A}^{\mu}\\
        &\hspace{0.5cm} - \frac{3 \, e^6 \, \text{Tr}\big(\mathcal{Y}_R^6\big)}{2} \, \frac{5+\xi}{6} \, \frac{1}{8} \, \overline{A}_{\mu} \overline{A}^{\mu} \overline{A}_{\nu} \overline{A}^{\nu}
        - e^4 \bigg( \frac{3\xi^2+519\xi+4558}{1440}\,\big(\mathcal{Y}_R^4\big)_{ji}\\
        &\hspace{0.5cm} - \frac{471\xi^2-92\xi+1221}{43200} \, \text{Tr}\big(\mathcal{Y}_R^2\big) \big(\mathcal{Y}_R^2\big)_{ji} \bigg) 
        \, \overline{\psi}_j \, i \, \overline{\slashed{\partial}} \, \mathbb{P}_{\mathrm{R}} \, \psi_i \bigg].
    \end{aligned}
\end{equation}


\end{appendices}

\printbibliography

\end{document}